\documentclass[lettersize,journal]{IEEEtran}
\usepackage[T1]{fontenc}   
\usepackage{amsmath,amsfonts}
\usepackage{array}
\usepackage[caption=false,font=normalsize,labelfont=sf,textfont=sf]{subfig}
\usepackage{textcomp}
\usepackage{stfloats}
\usepackage{url}
\usepackage{graphicx}
\usepackage{booktabs}
\usepackage{siunitx}
\usepackage{cite}
\graphicspath{{./}{./figures/}}
\begin{document}

\title{Directed Distance Fields for Constant-Time Ray Queries
on Gaussian Splatting}

\author{Subhankar Mishra%
\thanks{S. Mishra is with the School of Computer Sciences, National Institute of
Science Education and Research (NISER), 752050, India. 
E-mail: smishra@niser.ac.in.}} 

\markboth{IEEE Transactions on Visualization and Computer Graphics,~Vol.~XX, No.~X, 2026}%
{Mishra: Directed Distance Fields for Constant-Time Ray Queries on Gaussian Splatting}

\maketitle

\begin{abstract}
3D Gaussian Splatting (3DGS) renders new views of a scene in real time. Like every
rasterizer, it answers only primary rays, the rays from the camera through the
image. It cannot trace the secondary rays that shadows, ambient occlusion, and
global illumination need. We turn a trained 3DGS scene into a ray oracle by
distilling a Directed Distance Function (DDF). The DDF is a small neural field. It
takes a ray, given by an origin and a direction, and returns the distance to the
first surface and whether the ray hits anything. Each query is one forward pass. The
field is 52~MB, and its size does not depend on the number of Gaussians, so its cost
and memory stay flat as the scene grows. We make three points. First, we study what
supervision a DDF needs. Depth rendered from the Gaussians is too blurry to teach
thin parts, while clean distance supervision recovers them. Second, we measure
speed. The DDF is 26 to 72 times faster than sphere tracing an equivalent signed
distance field, and unlike a bounding volume hierarchy built over the Gaussians,
even on dedicated RT-core hardware, its query time and memory do not grow with the
scene. Third, we show a pipeline that needs no mesh: images give a 3DGS scene, a
neural surface gives clean distances, and the DDF learns from them. We use the DDF
as a secondary-ray oracle for global illumination. It reproduces reference
ray-traced shadows at 30.3~dB and ambient occlusion at 21.3~dB across 142 objects,
and on real captured scenes. Our codes are available at \url{https://github.com/smlab-niser/ddf-gs}.
\end{abstract}


\begin{figure*}[t]\centering
\includegraphics[width=\textwidth]{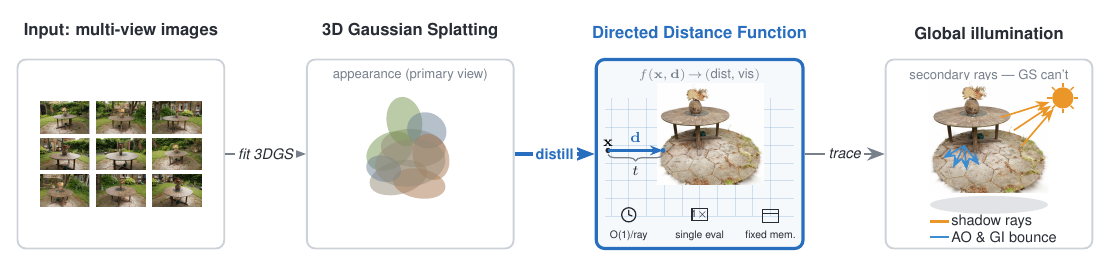}
\caption{We distill a Directed Distance Function (DDF) from a 3D Gaussian Splatting
scene. Multi-view images are fit to a 3DGS scene that holds appearance. We then
distill a DDF, a small field that returns, for any ray with origin $\mathbf{x}$ and
direction $\mathbf{d}$, the distance $t$ to the first surface and a visibility
value, in one network evaluation. A rasterizer gives only the primary view. The DDF
answers arbitrary secondary rays, so it can produce shadows, ambient occlusion, and
global illumination. It is 26 to 72 times faster than sphere tracing an equivalent
signed distance field, and its memory does not depend on scene size.}
\label{fig:teaser}
\end{figure*}

\section{Introduction}\label{sec:intro}
\IEEEPARstart{3}{D} Gaussian Splatting (3DGS)~\cite{kerbl2023gs} represents a scene
as a set of 3D Gaussians and rasterizes them into an image. It produces high
quality novel views and runs in real time. This has made it a common choice for
scene capture and display.

A rasterizer answers one kind of query. For each pixel it finds what the camera
sees along the ray through that pixel. These are primary rays. Many useful effects
need a different kind of query. To cast a shadow we ask whether a point on a surface
can see the light. To compute ambient occlusion we ask how much of the sky a point
can see. To trace one bounce of global illumination we ask what a surface point sees
in some direction. These are secondary rays. Their origins and directions are
arbitrary and are not known in advance. A rasterizer cannot answer them.

There are two ways to add ray queries to a Gaussian scene, and both have a cost.
The first is to trace the Gaussians directly. Recent
methods~\cite{moenneloccoz2024_3dgrt,raysplats2025} build a bounding volume
hierarchy (BVH) over a proxy shape for each Gaussian and traverse it on ray-tracing
hardware. The structure grows with the number of Gaussians, so its memory is $O(N)$
and its traversal cost rises as the scene gets larger. It also depends on
ray-tracing cores. The second way is to fit a signed distance field (SDF) to the
Gaussians and sphere trace it. Sphere tracing takes many small steps along each ray,
and each step is a full network evaluation, so the cost per ray is high.

We take a third way. We distill a Directed Distance Function (DDF) from the trained
Gaussian scene. A DDF maps a ray to the distance to the first surface along that ray
and to a visibility value. One forward pass answers one ray. The network is small
and fixed: 52~MB, with a size that does not depend on the number of Gaussians. So a
DDF turns a Gaussian scene into a ray oracle whose cost is constant per ray and
whose memory is constant in scene size. Directed distance fields have been studied
before, for shape representation~\cite{aumentado2022pddf} and for path
tracing~\cite{behera2023ddf}. They were not distilled from a Gaussian scene, and
they were not compared against sphere tracing an SDF, which is the natural baseline.

This paper studies the DDF as a ray oracle for Gaussian and implicit scenes. Our
contributions are:
\begin{itemize}
\item A DDF distilled from a Gaussian Splatting scene. It is a compact,
constant-time, constant-memory ray-intersection oracle that needs no ray-tracing
cores. It returns distance and visibility along any ray in one evaluation
(Sec.~\ref{sec:method}).
\item A study of what supervision a DDF needs to learn thin geometry. Depth rendered
from the Gaussians blurs thin parts and the DDF cannot learn them. Clean ray-cast
distance recovers them. The limit was the supervision, not the network
(Sec.~\ref{sec:supervision}).
\item A speed and scaling study. The DDF is 26 to 72 times faster than sphere
tracing an equivalent SDF (Table~\ref{tab:sdf}). Its query time and memory are flat
in scene size, while a BVH over the Gaussians grows in both (Table~\ref{tab:bvh},
Fig.~\ref{fig:scaling}).
\item A mesh-free oracle. Images give a Gaussian scene and a neural surface, and the
DDF learns from them with no mesh at any stage (Table~\ref{tab:e1}). We use it as a
global illumination oracle across 142 objects (Table~\ref{tab:e4}) and on real
captured scenes (Fig.~\ref{fig:realscene}).
\end{itemize}

\section{Related Work}\label{sec:related}

\textbf{Gaussian Splatting and ray tracing.}
3DGS~\cite{kerbl2023gs} fits a set of anisotropic 3D Gaussians to multi-view images
and rasterizes them for fast novel-view synthesis. Rasterization only serves the
camera view. To support arbitrary rays, recent work traces the Gaussians. 3D
Gaussian Ray Tracing~\cite{moenneloccoz2024_3dgrt} and
RaySplats~\cite{raysplats2025} build an acceleration structure over per-Gaussian
proxies and traverse it on ray-tracing cores. This gives correct intersections but
its memory and traversal cost grow with the number of Gaussians. Our DDF answers the
same intersection query in one evaluation, with a cost that does not depend on the
number of Gaussians.

\textbf{Neural implicit surfaces.}
A signed distance field stores the distance to the nearest surface at every
point~\cite{park2019deepsdf}. NeuS~\cite{wang2021neus} learns an SDF from images by
volume rendering. A multiresolution hash grid~\cite{mueller2022ingp} makes such
fields fast to query and train. An SDF gives a clean surface, but to find where a
ray meets that surface one must sphere trace it, which takes many evaluations per
ray. A DDF answers the same question in one evaluation because the distance it
stores is already measured along the query direction.

\textbf{Surface and SDF from Gaussian Splatting.}
Several methods recover geometry from a Gaussian scene. SuGaR~\cite{sugar2024} aligns
the Gaussians to a surface and extracts a mesh. 2D Gaussian
Splatting~\cite{2dgs2024huang} replaces the 3D Gaussians with flat disks for a more
accurate surface. GSDF~\cite{gsdf2024} fits a signed distance field next to the
Gaussians. These recover good geometry, but to answer a ray one must still trace the
mesh with an acceleration structure or sphere trace the SDF. The DDF answers the ray
directly in one evaluation.

\textbf{Directed distance fields.} 
A directed distance field stores the distance to the first surface as a function of
both a point and a direction. Probabilistic Directed Distance
Fields~\cite{aumentado2022pddf} use them for shape representation, and earlier work
from our group~\cite{behera2023ddf} uses a directional distance field for
path-traced rendering. We distill such a field from a Gaussian scene and study it as
a fast ray oracle.

\textbf{Visibility and illumination.}
NeRV~\cite{srinivasan2021nerv} learns a visibility field for relighting. A visibility
field answers whether a point can see a light. Recent work adds illumination to
Gaussian scenes by recovering materials and lighting. GS-IR~\cite{gsir2024} and
GaussianShader~\cite{gaussianshader2024} fit shading and reflectance to the
Gaussians, and Relightable 3D Gaussians~\cite{relightable3dgs2024} decompose the
scene for relighting and trace rays for shadows. These target appearance and
material. We instead provide a general ray oracle. The DDF returns the distance to
the surface along an arbitrary ray, so one field serves primary tracing, shadows,
ambient occlusion, and bounce queries, and global illumination is one application of
it.

\textbf{Data.}
We use objects from Google Scanned Objects~\cite{downs2022gso} and ShapeNet, and
real scenes from Mip-NeRF360~\cite{barron2022mipnerf360}.

\section{Method}\label{sec:method}
\begin{figure*}[t]\centering
\includegraphics[width=\textwidth]{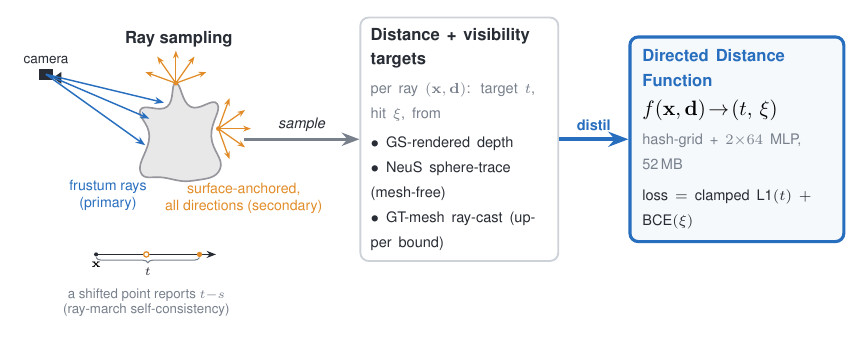}
\caption{How the DDF is supervised. We sample rays of two kinds: camera (frustum)
rays for the primary view, and surface-anchored rays in all directions for
secondary visibility. Along a hit ray, a point moved forward by $s$ must report the
remaining distance $t{-}s$ (ray-march self-consistency). The target distance $t$ and
hit $\xi$ for each ray come from one of three sources: depth rendered from the
Gaussians, a NeuS surface sphere-traced (no mesh), or a ground-truth mesh ray-cast
(an upper bound). We distil these targets into the DDF
$f(\mathbf{x},\mathbf{d})\!\to\!(t,\xi)$, a multiresolution hash grid with a small
$2{\times}64$ MLP (52~MB), with a clamped L1 loss on distance and a cross-entropy
loss on the hit.}
\label{fig:method} 
\end{figure*}

\subsection{The directed distance function}
Let a ray have origin $\mathbf{x}\in\mathbb{R}^3$ and unit direction
$\mathbf{d}\in\mathbb{S}^2$. The DDF is a function
\begin{equation}
f(\mathbf{x},\mathbf{d}) = (t,\;\xi),
\end{equation}
where $t$ is the distance from $\mathbf{x}$ to the first surface along $\mathbf{d}$,
and $\xi$ is a visibility value, the predicted probability that the ray meets the
scene at all. The hit point is $\mathbf{x}+t\,\mathbf{d}$ when $\xi$ is high.

We represent $f$ with a multiresolution hash grid that encodes the point
$\mathbf{x}$, following~\cite{mueller2022ingp}, together with a small encoding of the
direction $\mathbf{d}$. A two-layer MLP of width 64 reads this code and outputs $t$
and $\xi$. Almost all parameters live in the hash table; the MLP is tiny. The whole
field is 52~MB. This size is fixed. It does not grow with the number of Gaussians or
the size of the scene, which is the property that makes the oracle scale.

\subsection{Supervision}\label{sec:method-sup}
To train $f$ we need samples of the form (origin, direction, distance, hit). We get
them in three ways. We render depth from the Gaussians: a depth image gives, for
each camera ray, the distance to the surface. This is cheap and needs only the
Gaussian scene. We sphere trace a neural surface fit to the same images: this gives
clean distances and needs no mesh. We ray cast a ground-truth mesh: this is exact
and we use it as an upper bound and in ablations. Section~\ref{sec:supervision}
shows that the choice of source matters more than the network.

The set of rays also matters. Camera rays teach the field the view it will be
rendered from. Rays that start on the surface and point in all directions teach the
field the secondary visibility that shadows and ambient occlusion need. We sample
both. We also add a self-consistency term: along a ray that hits at distance $t$, a
point moved forward by $s$ should report distance $t-s$. Supervising these shifted
points keeps the field consistent along each ray and sharpens thin features. The
loss is an L1 term on distance, clamped to a maximum so far-away surfaces do not
dominate, plus a cross-entropy term on the hit flag.

We also tried to learn from the images directly, and it did not help. Adding a
volume-rendering color loss on top of the distance supervision, in the style of a
neural surface, did not bring back thin parts: most sampled rays land on the
background, and the path from color to the distance output is weak, so the extra loss
barely moved. Tracing the field step by step and supervising the trace from images
was unstable, because the trace is a sequential loop and its gradient gives no clean
signal for adding geometry the field has not already found. Clean distance
supervision stayed the reliable route.

\subsection{Querying the field}\label{sec:method-query}
Every query is one forward pass. To trace a primary ray we evaluate $f$ at the
camera and read $t$ and $\xi$. To test a shadow we take a surface point, offset it
slightly along its normal, and evaluate $f$ toward the light; if the ray is occluded
within the distance to the light, the point is in shadow. To compute ambient
occlusion we evaluate $f$ along a set of directions over the hemisphere and average
the hit flag. Each of these is a constant number of evaluations per ray. Sphere
tracing an SDF cannot do this, because each step depends on the previous one and a
ray needs many steps to converge. At inference we run the field in reduced precision
with a compiled forward pass. The speed numbers in Section~\ref{sec:speed} use this
setting.

\subsection{Implementation details}\label{sec:impl}
We encode the point $\mathbf{x}$ with a multiresolution hash
grid~\cite{mueller2022ingp} of 16 levels, two features per level, a table of $2^{19}$
entries per level, base resolution 16, and growth 1.5, over a box that bounds the
normalized object. We encode the direction $\mathbf{d}$ with four sinusoidal bands. A
two-layer MLP of width 64 reads the joined code and outputs $t$ and $\xi$. The hash
table holds about 13M parameters and the MLP about 8k, for a stored field of 52~MB. We
train with Adam for 30{,}000 steps, a batch of 16{,}384 rays, and a learning rate of
$3\times10^{-3}$ decayed by cosine to five percent of its value. The loss is an L1 term
on distance, clamped at the box scale, plus a cross-entropy term on the hit weighted at
0.3. We replace 30 percent of each batch with the ray-march shifted points of
Section~\ref{sec:method-sup}. Each object trains on one A100 in under an hour.

\section{Experiments and Results}\label{sec:exp}

\subsection{Setup}\label{sec:setup}
We fit a 3DGS scene for each object and scene. Objects come from Google Scanned
Objects~\cite{downs2022gso} and ShapeNet. Real scenes come from
Mip-NeRF360~\cite{barron2022mipnerf360}. We train each DDF on one A100 GPU. We
measure geometry with Chamfer distance and F-score, global illumination with PSNR
against a reference ray tracer (embree), and speed with wall-clock time using GPU
synchronization.

\subsection{What supervision does a DDF need?}\label{sec:supervision}
Earlier DDFs trained from Gaussian-rendered depth recovered the rough shape of an
object but lost thin parts, such as the legs of an animal figure or the handle of a
mug. We tested whether this was a limit of the network or of the supervision. We
trained the same network from clean distances obtained by ray casting the ground-truth
mesh. A leg only a few pixels wide is averaged away when depth is rendered from soft
Gaussians, so the depth image carries no sharp signal there. A ray cast against the
surface measures the distance exactly.

Table~\ref{tab:sup} reports Chamfer distance over ten objects for three supervision
sources. Clean distance lowers the mean from 0.082 to 0.065, a drop of 21 percent, and
the gains sit on the thin-feature objects: the mug handle (0.194 to 0.067), the chair
legs (0.120 to 0.054), and the turtle limbs (0.103 to 0.067). On compact objects whose
shape Gaussian depth already captures, clean distance neither helps nor hurts much. The
network did not change between these columns; only the supervision did. The same
clean-distance DDF also matches the neural surface (NeuS) it could have learned from,
0.065 against 0.076 in the mean, and beats it where NeuS over-extrudes thin parts, such
as the airplane wings (0.043 against 0.151) and the chair legs (0.054 against 0.152).
NeuS stays ahead on the compact figurines. On two objects, spino and bottle, clean
distance does not help, because the sphere-traced point cloud there is noisier and the
surface reconstruction fattens it, the ray-primitive behavior we return to in
Section~\ref{sec:discussion}.

\subsection{The encoding carries the geometry}\label{sec:encoding}
Our first DDF used a sinusoidal encoding with a wide MLP. We replace the encoding with
a multiresolution hash grid and shrink the MLP to two layers. Table~\ref{tab:enc}
ablates this on the bull, with the supervision and the extraction held fixed. The hash
grid lowers Chamfer distance from 0.187 to 0.117, a drop of 37 percent, and its median
falls by more than half, while the MLP shrinks from 397k to 8k parameters. Shrinking
the sinusoidal MLP instead only makes it worse. The encoding, not the size of the MLP,
carries the geometry.

\subsection{Speed and scaling}\label{sec:speed}
We first compare the DDF against sphere tracing an equivalent neural surface. Both
answer the same question, where a ray meets the surface. We use networks of similar
size and the same hardware and rays. Table~\ref{tab:sdf} reports the result. The DDF
is 26 to 72 times faster across ray counts and across convergence thresholds. The
reason is structural: sphere tracing is a sequential loop, and each iteration is a
full evaluation, while the DDF answers in one pass.

\begin{figure}[t]\centering
\includegraphics[width=\columnwidth]{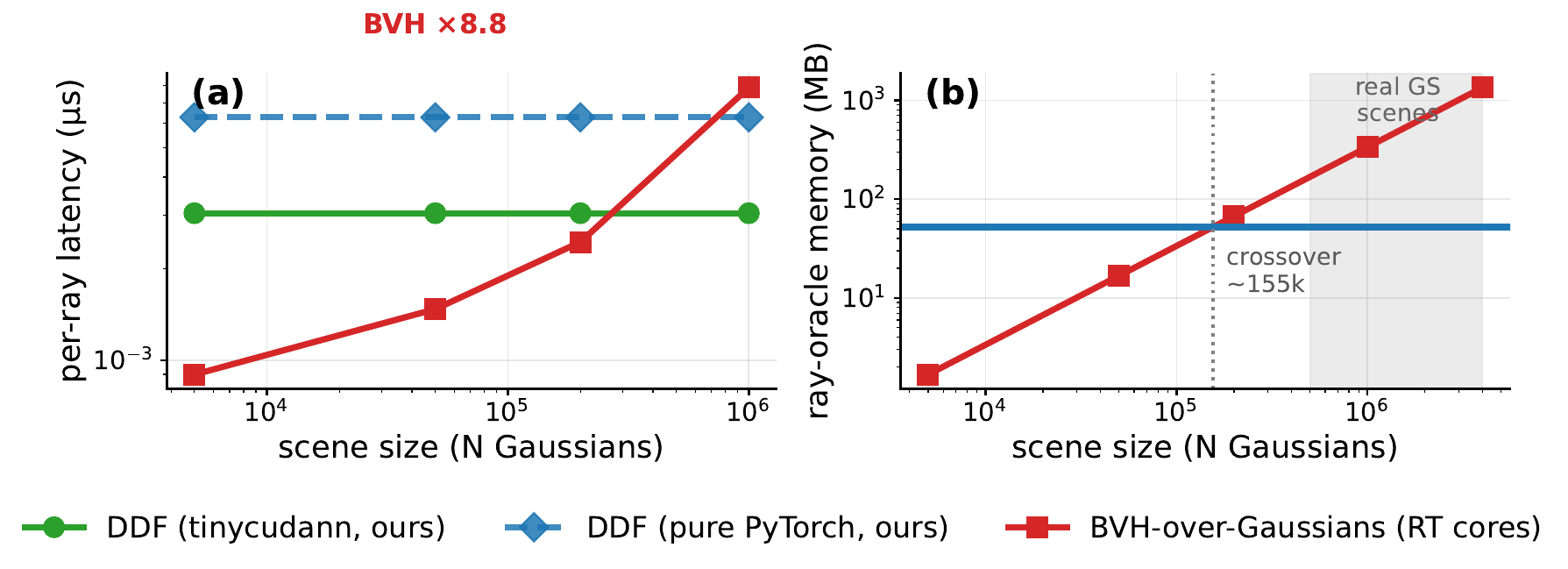}
\caption{Scaling against scene size. The BVH uses NVIDIA OptiX on an RTX 2080 Ti
(dedicated RT cores); the DDF is shown in two forms, pure-PyTorch (dashed) and the
same architecture re-built with CUDA-fused kernels via tinycudann (solid). (a)
Per-ray query latency: both DDFs are flat, because the cost never sees the Gaussian
count, while the BVH over Gaussians rises, by about nine times from 5K to 1M
Gaussians (shown at a 1M-ray frame; the rising trend holds at all ray counts,
Table~\ref{tab:bvh}). The BVH crosses both DDF lines well below 1M Gaussians. (b)
Ray-oracle memory: the DDF is a fixed 52~MB, while the BVH proxy geometry grows at
336~B per Gaussian and crosses 52~MB near 155K Gaussians. Real scenes have 0.5 to
4M Gaussians (shaded).}
\label{fig:scaling} 
\end{figure}

We then compare against a BVH built over the Gaussians, which is the structure that
ray-traced Gaussian methods rely on. Table~\ref{tab:bvh} and Fig.~\ref{fig:scaling}
show the trend. The DDF cost is flat in the number of Gaussians, because the network
never sees that number. The BVH cost rises with it. The memory gap is sharp: the DDF
is 52~MB at any scene size, while the proxy geometry of the BVH grows at 336 bytes
per Gaussian and passes 52~MB near 155k Gaussians. Real scenes have hundreds of
thousands to millions of Gaussians, so at scene scale the DDF is the smaller and
flatter oracle.

We use the strongest BVH baseline we can: NVIDIA OptiX (via Mitsuba~3) on an RTX
2080 Ti, which routes ray-triangle tests through dedicated RT cores. This is the
underlying ray-intersection primitive that 3DGRT~\cite{moenneloccoz2024_3dgrt} and
RaySplats~\cite{raysplats2025} use; our measurement therefore characterizes the
ray-query cost shared by all BVH-over-Gaussians methods, isolated from the radiance
integration they additionally perform per ray. A CPU Embree implementation of the
same BVH is 50 to 1000 times slower per ray. On RT-core
hardware the DDF and the BVH are no longer orders of magnitude apart, but the
structural claims hold. The DDF cost stays flat in the number of Gaussians; the BVH
cost still rises with it, by about nine times from 5K to 1M Gaussians at 1M rays,
even with hardware acceleration. The DDF memory stays fixed; the BVH memory still
grows.

For absolute per-ray latency the picture is nuanced. On small scenes (a few thousand
Gaussians) the RT-core BVH is several times faster per ray than the DDF, because
hardware ray-triangle traversal is highly optimized. As the scene grows, the BVH
latency rises while the DDF stays flat; at the scene scales real GS captures use
(hundreds of thousands to millions of Gaussians) the two methods become competitive.
We also report the DDF with CUDA-fused kernels (tinycudann)~\cite{mueller2022ingp}.
This is a drop-in engineering replacement of the same hash-grid plus MLP
architecture, not a different method. It roughly halves the DDF cost at large ray
counts and pushes the DDF clearly ahead of the RT-core BVH at scene scale (at 1M
Gaussians and 100K rays, 0.005 vs.\ 0.009 microseconds per ray). The DDF's enduring
advantages are the constant cost in scene size, the constant memory, the lack of
dependence on RT-core hardware, and a single forward pass per ray instead of an
iterative traversal. The DDF was measured on an A100 and the RT-core BVH on an RTX
2080 Ti; the A100 has no RT cores. The same DDF on RTX hardware would be modestly
slower than reported, but the scaling and memory claims do not depend on the GPU
choice.

We also measure the DDF on a real captured scene, the Tanks and Temples
\emph{train} scene with 661,693 Gaussians~\cite{knapitsch2017tnt}. Here we compare
against re-rasterizing the scene with gsplat, which is the cost of answering the
same rays by rendering. For sparse queries the DDF is far faster: 641 times at
1K rays and 603 times at 10K rays. At 100K rays it is 42.6 times faster
(0.50~ms vs.\ 21.1~ms). At a full frame of 1M rays the two are on par
(4.81~ms vs.\ 4.45~ms), because rasterization is then bound by writing the frame
buffer. So at real scene scale the DDF wins for the sparse and moderate ray counts
that secondary-ray queries use, and only ties rasterization at a dense full frame.

\subsection{A mesh-free oracle}\label{sec:meshfree}
A DDF trained from a ground-truth mesh is a clean test of the network, but it needs a
mesh. Real scenes have none. We therefore train the secondary-ray oracle from a
neural surface fit to the images, with no mesh at any stage. Table~\ref{tab:e1}
reports shadow-ray agreement against a reference mesh ray tracer. The mesh-free
oracle reaches 75.5\% on average. This is far above the frustum-only baseline at
50.2\%, and it approaches the mesh-trained upper bound at 94.8\%. The gap that
remains comes from the neural surface, which oversmooths thin parts, and not from
the DDF. A sharper surface would close it.

\subsection{Global illumination}\label{sec:gi}
\begin{figure}[t]\centering
{\scriptsize\makebox[0.3333\columnwidth][c]{GS only}%
\makebox[0.3333\columnwidth][c]{GS + DDF GI (ours)}%
\makebox[0.3333\columnwidth][c]{GS + embree GI (GT)}}\\[1pt]
\includegraphics[width=\columnwidth]{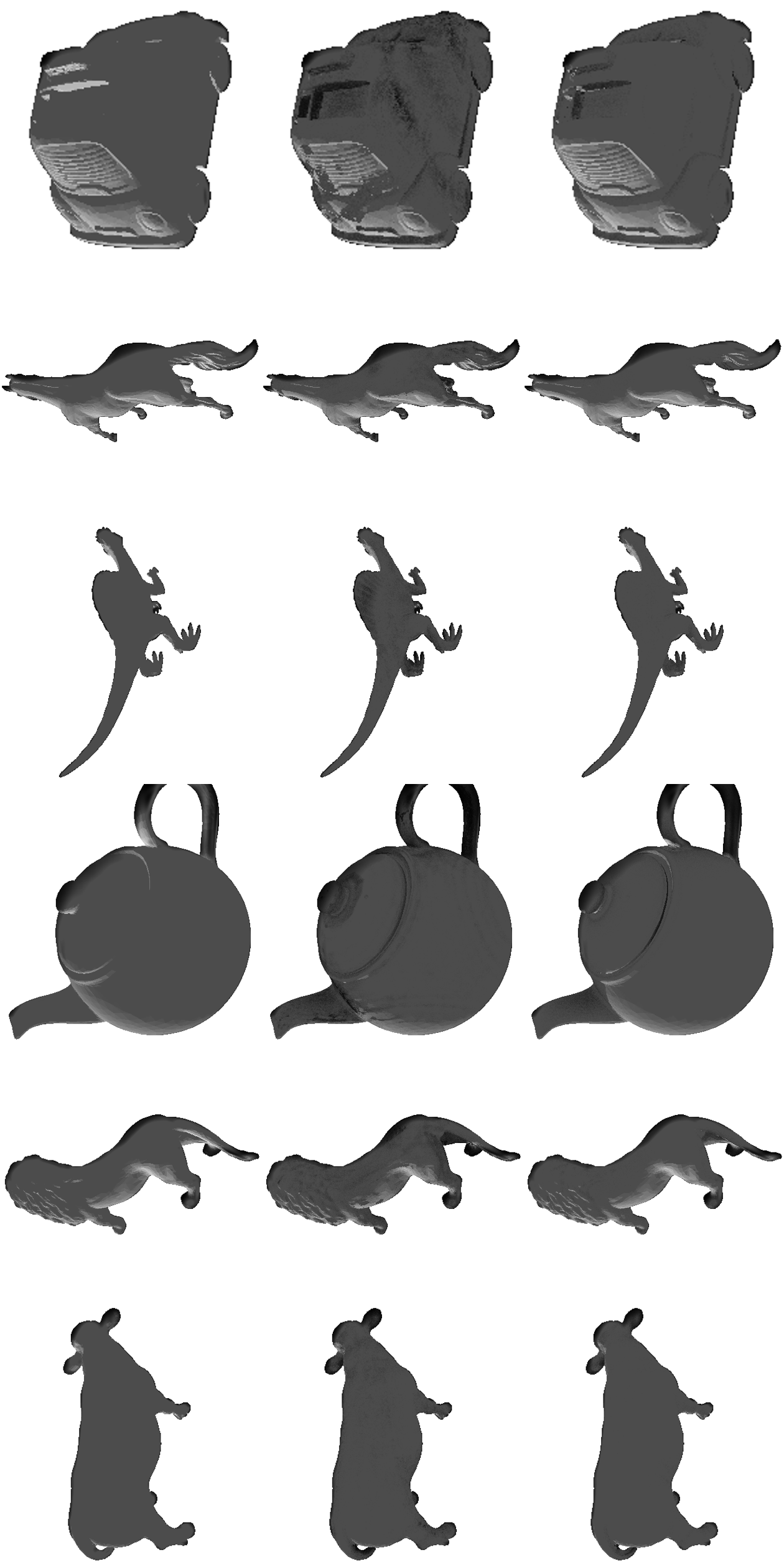}
\caption{DDF-traced global illumination on objects (rows). Left: the Gaussian
render with no GI. Middle: our DDF self-shadow and ambient occlusion. Right: the
embree ground truth. The per-ray DDF visibility matches the reference ray tracer.}
\label{fig:hero}
\end{figure}

\begin{figure*}[t]\centering
{\footnotesize\makebox[0.2867\textwidth][c]{GS only}%
\makebox[0.2867\textwidth][c]{DDF GI map}%
\makebox[0.2867\textwidth][c]{GS + DDF GI}}\\[2pt]
\includegraphics[width=0.86\textwidth]{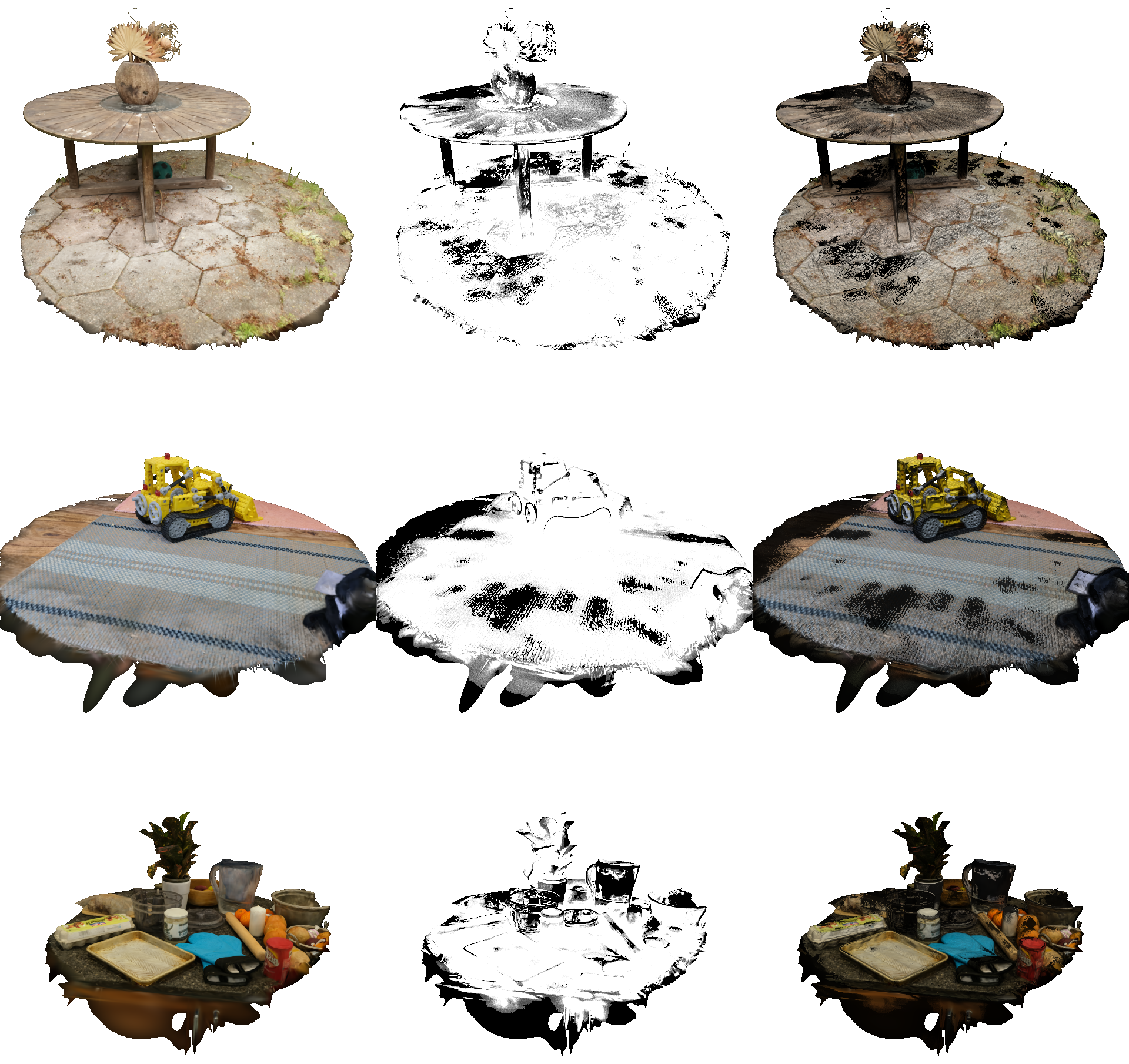}
\caption{Mesh-free global illumination on real captured Mip-NeRF360 scenes (rows).
Left: the Gaussian render, which holds appearance. Middle: the DDF visibility map,
where dark means occluded, computed at one evaluation per ray. Right: the Gaussian
scene with DDF-traced shadows and ambient occlusion. No mesh is used at any stage.}
\label{fig:realscene}
\end{figure*}

We use the DDF as the secondary-ray oracle in a renderer and compare its output to a
reference ray tracer on the same geometry. Only the oracle differs, so any
difference is the oracle. Table~\ref{tab:e4} reports the result over 142 objects; the
per-object numbers are in the supplemental material. The DDF reproduces the reference
shadows at 30.3~dB and ambient occlusion at 21.3~dB. Fig.~\ref{fig:hero} shows the effect on objects, and
Fig.~\ref{fig:realscene} shows it on real captured scenes. The Gaussian render gives
appearance but no shadows. The DDF adds shadows and ambient occlusion that the
Gaussians alone cannot produce, with no mesh in the pipeline.


\subsection{Positioning}\label{sec:position}
Table~\ref{tab:position} places the DDF against recent ray-query methods for
Gaussian and implicit scenes. It is the only one that answers in constant time, uses
constant memory, needs no ray-tracing cores, and returns distance along an arbitrary
ray. Methods that trace the Gaussians grow in time and memory with the scene.
Methods that fit an SDF need several evaluations per ray. A visibility field answers
only whether a light is seen, not the distance along a general ray.


\begin{table}[t]\centering
\caption{\textbf{Supervision sets the quality, not the network.} Chamfer distance
(mean / median; sphere-trace extraction) for ten objects with the same hash-grid
DDF, varying only the supervision source. NeuS is shown for reference and needs
sphere tracing to query. Lower is better; best in each row in bold.}
\label{tab:sup}
\setlength{\tabcolsep}{4pt}
\begin{tabular}{l ccc}
\toprule
object & GS depth & clean distance & NeuS \\
\midrule
bull     & 0.072 / 0.054 & 0.061 / 0.048 & \textbf{0.022 / 0.020} \\
lion     & 0.079 / 0.059 & 0.076 / 0.060 & \textbf{0.048 / 0.044} \\
spino    & \textbf{0.037 / 0.029} & 0.103 / 0.077 & 0.043 / 0.034 \\
mug      & 0.194 / 0.129 & \textbf{0.067 / 0.047} & 0.091 / 0.062 \\
turtle   & 0.103 / 0.077 & 0.067 / 0.056 & \textbf{0.053 / 0.048} \\
airplane & \textbf{0.036 / 0.028} & 0.043 / 0.030 & 0.151 / 0.119 \\
chair    & 0.120 / 0.036 & \textbf{0.054 / 0.044} & 0.152 / 0.022 \\
car      & 0.063 / 0.048 & \textbf{0.056 / 0.041} & 0.074 / 0.030 \\
bottle   & 0.057 / 0.050 & 0.069 / 0.069 & \textbf{0.046 / 0.015} \\
sofa     & 0.064 / 0.050 & \textbf{0.056 / 0.039} & 0.081 / 0.035 \\
\midrule
mean     & 0.082 & \textbf{0.065} & 0.076 \\
\bottomrule
\end{tabular}
\end{table}

\begin{table}[t]\centering
\caption{\textbf{Encoding ablation on the bull.} Same depth supervision and the same
extraction (UDF to marching cubes at iso 0.05); only the input encoding changes. The
extraction differs from Table~\ref{tab:sup}'s sphere tracing, so the bull values are
not directly comparable across the two tables. Lower is better.}
\label{tab:enc}
\begin{tabular}{l c cc}
\toprule
encoding & MLP params & CD mean & CD median \\
\midrule
sinusoidal MLP, $256{\times}6$ & 397k & 0.187 & 0.169 \\
sinusoidal MLP, $128{\times}4$ & 84k  & 0.222 & 0.192 \\
\textbf{hash grid $+$ $2{\times}64$ MLP} & \textbf{8k} & \textbf{0.117} & \textbf{0.063} \\
\bottomrule
\end{tabular}
\end{table}

\begin{table}[t]\centering
\caption{\textbf{Per-ray query latency, DDF vs.\ BVH-over-Gaussians} ($\mu$s per ray).
DDF runs on an A100; the RT-core BVH uses NVIDIA OptiX on an RTX 2080 Ti; the CPU
BVH uses Embree on x86. The same DDF architecture is reported with our pure-PyTorch
implementation and with CUDA-fused kernels (tinycudann). Lower is better.}
\label{tab:bvh}
\setlength{\tabcolsep}{4pt}
\small
\begin{tabular}{l rrrr}
\toprule
& \multicolumn{4}{c}{number of rays} \\
\cmidrule(lr){2-5}
scene size & 1K & 10K & 100K & 1M \\
\midrule
\textbf{DDF, pure-PyTorch (any $N_{\mathrm{g}}$)} & 0.264 & 0.028 & 0.0083 & 0.0063 \\
\textbf{DDF, tinycudann (any $N_{\mathrm{g}}$)} & 0.448 & 0.043 & \textbf{0.0050} & \textbf{0.0030} \\
\midrule
\multicolumn{5}{l}{\emph{BVH on RT cores (RTX 2080 Ti, OptiX)}} \\
BVH @ 5K   & 0.118 & 0.012 & 0.002 & 0.0009 \\
BVH @ 50K  & 0.099 & 0.012 & 0.002 & 0.0015 \\
BVH @ 200K & 0.110 & 0.014 & 0.004 & 0.0024 \\
BVH @ 1M   & 0.117 & 0.020 & 0.009 & 0.0079 \\
\midrule
\multicolumn{5}{l}{\emph{BVH on CPU (Embree)}} \\
BVH @ 5K   & 0.868 & 1.019 & 1.028 & 0.943 \\
BVH @ 50K  & 1.413 & 1.666 & 1.603 & 1.521 \\
BVH @ 200K & 2.256 & 3.151 & 2.643 & 2.552 \\
BVH @ 1M   & 4.954 & 8.079 & 6.959 & 6.529 \\
\bottomrule
\end{tabular}
\end{table}

\begin{table}[t]\centering
\caption{\textbf{DDF (one eval/ray) vs.\ sphere-tracing an equivalent NeuS SDF}
(same GPU, same rays, comparable network size). Sphere-tracing is sequential; the
DDF answers in a single forward pass.}
\label{tab:sdf}
\begin{tabular}{r rr r}
\toprule
rays & DDF (ms) & SDF sphere-trace (ms) & speedup \\
\midrule
1K   & 0.17 & 68.2  & $403\times$ \\
10K  & 1.07 & 85.0  & $79\times$ \\
100K & 0.74 & 109.5 & $147\times$ \\
1M   & 6.08 & 454.2 & $75\times$ \\
\bottomrule
\end{tabular}
\end{table}

\begin{table}[t]\centering
\caption{\textbf{Mesh-free secondary-ray oracle.} Shadow-ray agreement vs.\
GT-mesh embree (\%). \emph{NeuS-omnidir} (ours; images$\to$NeuS$\to$DDF, no mesh)
approaches the mesh-trained upper bound and far exceeds the frustum-only baseline.}
\label{tab:e1}
\begin{tabular}{l c ccc}
\toprule
object & shadow rate & gtmesh (UB) & \textbf{NeuS-omnidir} & neus\_v3 \\
\midrule
bull   & 49.1 & 94.3 & \textbf{79.4} & 39.5 \\
mug    & 65.9 & 97.5 & \textbf{79.5} & 63.3 \\
turtle & 56.3 & 92.7 & \textbf{69.4} & 53.2 \\
lion   & 45.8 & 94.9 & \textbf{73.6} & 44.7 \\
\midrule
\textbf{mean} & n/a & \textbf{94.8} & \textbf{75.5} & \textbf{50.2} \\
\bottomrule
\end{tabular}
\end{table}

\begin{table}[t]\centering
\caption{\textbf{GI fidelity at scale.} The O(1) DDF oracle reproduces embree's
shadows/AO across 142 objects (PSNR vs.\ GT-mesh embree, dB; mean/median).}
\label{tab:e4}
\begin{tabular}{l c cc c}
\toprule
subset & $n$ & shadow PSNR & AO PSNR & AO MAE \\
\midrule
all      & 142 & 30.3 / 29.5 & 21.3 / 20.8 & 0.216 \\
GSO      & 29  & 30.8 / 30.7 & 21.3 / 21.7 & 0.175 \\
ShapeNet & 113 & 30.2 / 29.3 & 21.3 / 20.4 & 0.227 \\
\bottomrule
\end{tabular}
\end{table}

\begin{table*}[t]\centering
\caption{\textbf{Positioning} vs.\ recent ray-query methods for GS and implicit
scenes. The DDF is the only constant-time, constant-memory, RT-core-free oracle
that predicts distance along an arbitrary ray and is representation-agnostic.}
\label{tab:position}
\begin{tabular}{l c c c c c}
\toprule
method & query per ray & memory & grows with $N_{\mathrm{g}}$? & no RT cores & predicts \\
\midrule
\textbf{DDF (ours)} & \textbf{O(1), one eval} & \textbf{constant} & \textbf{no} & \textbf{yes} & \textbf{distance + visibility, any ray} \\
3DGRT / RaySplats   & O($\log N$) traverse    & grows, O($N$)     & yes         & no          & radiance at the per-Gaussian hit \\
SDF-from-GS         & O($k$) sphere-trace     & constant          & no          & yes         & distance (omnidirectional SDF) \\
NeRV / visibility field & O(1) eval           & constant          & no          & yes         & light visibility only \\
\bottomrule
\end{tabular}
\end{table*}

\section{Discussion, Limitations, and Future Work}\label{sec:discussion}
The global illumination from a mesh-free real scene is real but modest. The reason
is geometry. Without a mesh we read surface normals from the rendered depth, and
these normals are soft, so the shading is gentle rather than sharp. We therefore show
the DDF visibility directly as a map, where the per-ray result is clear, rather than
hide it inside a soft shaded image.

Our BVH baseline runs on a CPU, so it does not match a tuned ray-tracing-core kernel
in absolute time. We state this and rely on the parts of the comparison that do not
depend on hardware, namely the flat cost and the flat memory of the DDF as the scene
grows.

The mesh-free oracle is bounded by the neural surface it learns from. When that
surface oversmooths thin parts, the oracle inherits the error. A sharper image-based
surface would raise the 75.5\% figure toward the mesh-trained 94.8\%.

Finally, a DDF is a ray-query primitive, not a smooth-surface generator. It is best
used where a mesh is not wanted, such as an implicit pipeline, a differentiable
query for inverse rendering, or a moving scene.

Two further limits are worth naming. Ambient occlusion is weaker than shadows,
21.3~dB against 30.3~dB, because it averages many rays and the small per-ray error of
the field adds up. The oracle is also accurate only for the ray origins it was
trained on, which lie on or near the surface. Rays that start far from any surface,
such as a point on a ground plane below the object, fall outside this set, and there
the field reports occlusion too often. Training with such origins would extend the
oracle to them.

\textbf{Future work.} A sharper image-based surface would raise the mesh-free oracle
toward the mesh-trained bound and improve the real-scene result. Supervising rays
from ground and volumetric origins would let the same field cast contact shadows on a
floor and support full-scene, not only on-surface, global illumination. A direct
comparison against a tuned ray-tracing-core kernel would complete the speed picture
at small scene scale, where that hardware is strongest. The constant per-ray cost
also fits multi-bounce illumination and moving scenes, where a fixed acceleration
structure must be rebuilt as the scene changes.

\section{Conclusion}\label{sec:conclusion}
We distilled a Directed Distance Function from a Gaussian Splatting scene and used it
as a secondary-ray oracle. It answers any ray in one evaluation. Its cost and memory
do not grow with the scene. It is 26 to 72 times faster than sphere tracing an
equivalent surface, and it produces shadows and ambient occlusion that a rasterizer
cannot. The supervision, not the network, sets the quality, and clean distance
supervision recovers thin geometry. A pipeline that starts from images and uses no
mesh closes the loop. The result is a small, constant-time ray oracle for Gaussian
and implicit scenes whose advantage over a bounding volume hierarchy grows as the
scene grows.

\section*{Acknowledgments}
This work was supported by the Department of Atomic Energy, Government of India,
under grant RIN4009.

\bibliographystyle{IEEEtran}
\bibliography{refs}

@article{knapitsch2017tnt,
  title={Tanks and Temples: Benchmarking Large-Scale Scene Reconstruction},
  author={Knapitsch, Arno and Park, Jaesik and Zhou, Qian-Yi and Koltun, Vladlen},
  journal={ACM Transactions on Graphics}, volume={36}, number={4}, year={2017},
  doi={10.1145/3072959.3073599}}

@article{kerbl2023gs,
  title={3D Gaussian Splatting for Real-Time Radiance Field Rendering},
  author={Kerbl, Bernhard and Kopanas, Georgios and Leimk{\"u}hler, Thomas and Drettakis, George},
  journal={ACM Transactions on Graphics}, volume={42}, number={4}, articleno={139},
  pages={1--14}, year={2023}, doi={10.1145/3592433}}

@inproceedings{wang2021neus,
  title={NeuS: Learning Neural Implicit Surfaces by Volume Rendering for Multi-view Reconstruction},
  author={Wang, Peng and Liu, Lingjie and Liu, Yuan and Theobalt, Christian and Komura, Taku and Wang, Wenping},
  booktitle={Advances in Neural Information Processing Systems (NeurIPS)}, year={2021}}

@article{mueller2022ingp,
  title={Instant Neural Graphics Primitives with a Multiresolution Hash Encoding},
  author={M{\"u}ller, Thomas and Evans, Alex and Schied, Christoph and Keller, Alexander},
  journal={ACM Transactions on Graphics}, volume={41}, number={4}, articleno={102},
  pages={1--15}, year={2022}, doi={10.1145/3528223.3530127}}

@inproceedings{park2019deepsdf,
  title={DeepSDF: Learning Continuous Signed Distance Functions for Shape Representation},
  author={Park, Jeong Joon and Florence, Peter and Straub, Julian and Newcombe, Richard and Lovegrove, Steven},
  booktitle={IEEE/CVF Conf.\ on Computer Vision and Pattern Recognition (CVPR)}, pages={165--174}, year={2019}}

@inproceedings{aumentado2022pddf,
  title={Representing 3D Shapes with Probabilistic Directed Distance Fields},
  author={Aumentado-Armstrong, Tristan and Tsogkas, Stavros and Dickinson, Sven and Jepson, Allan},
  booktitle={IEEE/CVF Conf.\ on Computer Vision and Pattern Recognition (CVPR)}, pages={19343--19354}, year={2022}}

@inproceedings{behera2023ddf,
  title={Neural directional distance field object representation for uni-directional path-traced rendering},
  author={Behera, Annada Prasad and Mishra, Subhankar},
  booktitle={2023 14th International Conference on Computing Communication and Networking Technologies (ICCCNT)},
  pages={1--6},
  year={2023},
  organization={IEEE}
}

@inproceedings{srinivasan2021nerv,
  title={NeRV: Neural Reflectance and Visibility Fields for Relighting and View Synthesis},
  author={Srinivasan, Pratul P. and Deng, Boyang and Zhang, Xiuming and Tancik, Matthew and Mildenhall, Ben and Barron, Jonathan T.},
  booktitle={IEEE/CVF Conf.\ on Computer Vision and Pattern Recognition (CVPR)}, pages={7495--7504}, year={2021}}

@inproceedings{downs2022gso,
  title={Google Scanned Objects: A High-Quality Dataset of 3D Scanned Household Items},
  author={Downs, Laura and Francis, Anthony and Koenig, Nate and Kinman, Brandon and Hickman, Ryan and Reymann, Krista and McHugh, Thomas B. and Vanhoucke, Vincent},
  booktitle={IEEE Int.\ Conf.\ on Robotics and Automation (ICRA)}, pages={2553--2560}, year={2022}}

@inproceedings{barron2022mipnerf360,
  title={Mip-NeRF 360: Unbounded Anti-Aliased Neural Radiance Fields},
  author={Barron, Jonathan T. and Mildenhall, Ben and Verbin, Dor and Srinivasan, Pratul P. and Hedman, Peter},
  booktitle={IEEE/CVF Conf.\ on Computer Vision and Pattern Recognition (CVPR)}, pages={5470--5479}, year={2022}}

@article{moenneloccoz2024_3dgrt,
author = {Moenne-Loccoz, Nicolas and Mirzaei, Ashkan and Perel, Or and de Lutio, Riccardo and Martinez Esturo, Janick and State, Gavriel and Fidler, Sanja and Sharp, Nicholas and Gojcic, Zan},
title = {3D Gaussian Ray Tracing: Fast Tracing of Particle Scenes},
year = {2024},
issue_date = {December 2024},
publisher = {Association for Computing Machinery},
address = {New York, NY, USA},
volume = {43},
number = {6},
issn = {0730-0301},
url = {https://doi.org/10.1145/3687934},
doi = {10.1145/3687934},
journal = {ACM Trans. Graph.},
month = nov,
articleno = {232},
numpages = {19}
}

@article{raysplats2025,
  title={RaySplats: Ray Tracing based Gaussian Splatting},
  author={Byrski, Krzysztof and Mazur, Marcin and Tabor, Jacek and Dziarmaga, Tadeusz and K{\k{a}}dzio{\l}ka, Marcin and Baran, Dawid and Spurek, Przemys{\l}aw},
  journal={arXiv preprint arXiv:2501.19196},
  year={2025}
}

@inproceedings{sugar2024,
  title={SuGaR: Surface-Aligned Gaussian Splatting for Efficient 3D Mesh Reconstruction and High-Quality Mesh Rendering},
  author={Gu{\'e}don, Antoine and Lepetit, Vincent},
  booktitle={IEEE/CVF Conf.\ on Computer Vision and Pattern Recognition (CVPR)},
  pages={5354--5363}, year={2024}, doi={10.1109/CVPR52733.2024.00512}}

@inproceedings{2dgs2024huang,
author = {Huang, Binbin and Yu, Zehao and Chen, Anpei and Geiger, Andreas and Gao, Shenghua},
title = {2D Gaussian Splatting for Geometrically Accurate Radiance Fields},
year = {2024},
isbn = {9798400705250},
publisher = {Association for Computing Machinery},
address = {New York, NY, USA},
url = {https://doi.org/10.1145/3641519.3657428},
doi = {10.1145/3641519.3657428},
booktitle = {ACM SIGGRAPH 2024 Conference Papers},
articleno = {32},
numpages = {11},
location = {Denver, CO, USA},
series = {SIGGRAPH '24}
}

@article{gsdf2024,
title={GSDF: 3DGS Meets SDF for Improved Neural Rendering and Reconstruction},
  author={Yu, Mulin and Lu, Tao and Xu, Linning and Jiang, Lihan and Xiangli, Yuanbo and Dai, Bo},
  journal={Advances in Neural Information Processing Systems},
  volume={37},
  pages={129507--129530},
  year={2024}
}

@inproceedings{gsir2024,
  title={GS-IR: 3D Gaussian Splatting for Inverse Rendering},
  author={Liang, Zhihao and Zhang, Qi and Feng, Ying and Shan, Ying and Jia, Kui},
  booktitle={Proceedings of the IEEE/CVF Conference on Computer Vision and Pattern Recognition},
  pages={21644--21653},
  year={2024}
}

@inproceedings{gaussianshader2024,
  title={GaussianShader: 3D Gaussian Splatting with Shading Functions for Reflective Surfaces},
  author={Jiang, Yingwenqi and Tu, Jiadong and Liu, Yuan and Gao, Xifeng and Long, Xiaoxiao and Wang, Wenping and Ma, Yuexin},
  booktitle={IEEE/CVF Conf.\ on Computer Vision and Pattern Recognition (CVPR)},
  pages={5322--5332}, year={2024}, doi={10.1109/CVPR52733.2024.00509}}

@inproceedings{relightable3dgs2024,
  title={Relightable 3D Gaussians: Realistic Point Cloud Relighting with BRDF Decomposition and Ray Tracing},
  author={Gao, Jian and Gu, Chun and Lin, Youtian and Zhu, Hao and Cao, Xun and Zhang, Li and Yao, Yao},
  booktitle={European Conf.\ on Computer Vision (ECCV)}, pages={73--89}, year={2024},
  doi={10.1007/978-3-031-72995-9_5}}

\end{document}